\documentclass{article}[12pt]
\usepackage{latexsym}
\begin{document}

\title{Stellar explosion in the weak field approximation of the Brans-Dicke theory}
\author{Hamity Victor H and Barraco Daniel E \\ Fa.M.A.F., Universidad Nacional de C\'ordoba\\
Ciudad Universitaria\\ C\'ordoba 5000, Argentina}
\date{June 30, 2005}
\maketitle
\begin{abstract}
We treat a very crude model of an exploding star, in the weak
field approximation of the Brans-Dicke theory, in a scenario that
resembles some characteristics data of a Type Ia Supernova. The
most noticeable feature, in the electromagnetic component, is the
relationship  between the absolute magnitude at maximum brightness
of the star and the decline rate in one magnitude from that
maximum. This characteristic has become one of the most accurate
method to measure luminosity distances to objects at cosmological
distances\cite{phillips, supern}. An interesting result is that
the
 active mass associated with the scalar field is totally
radiated to infinity,  representing a mass loss in the ratio of the ``tensor"
component to the scalar component of 1 to $(2 \, \omega + 3)$ ($\omega$ is the Brans-Dicke parameter), in
agreement with a general result of Hawking \cite{hawking}. Then,
this model shows explicitly, in a dynamical case, the mechanism of
radiation  of scalar field, which is necessary to understand the
Hawking result.
\\
\\
PACS numbers(s): 04.30.Db, 04.50.+h,04.80.Cc
\end{abstract}

\section{Introduction}

Although the problem of constructing an exact solution in the
Brans-Dicke (BD) theory, that may be considered a generalization
of  Vaidya solution\cite{vaidya},  has been addressed in the
past\cite{antecedentes}, this solution has not been found yet.
That generalization  must reduce to Vaidya solution corresponding
to the external gravitational field of a radially radiating star,
in the limit of GR, and also, in the limit that the radiation flux
vanishes, we must be able to obtain one of the empty space
solution of Brans\cite{brans}.

The interest of having a solution in the BD theory associated with
electromagnetic radiation is to use it in the construction of a
model of an exploding star that includes the ejection of a shell
of mater, electromagnetic  radiation in a similar way as  in a
model in GR\cite{hamity2,lake,hs}, and scalar field radiation.
This may be in consonance with a star model in which a highly
evolved core implodes to nuclear densities while at the same time
the outer most layers of matter are blown at high speed. This idea
can be realized in the weak field approximation of BD theory using
the method of Barros and Romero\cite{barros} to construct a
generalization of Vaidya solution that it is matched through a
spherical singular hypersurface to a  satisfactory interior
solution.

The formalism of a singular hypersurface in  BD theory has been
developed and used in the description of bubbles dynamics in
extended inflation models of the Universe\cite{sakai,dalia}, among
others applications. While Barrabes et al\cite{barrabes} treat  a
singular hypersurface in BD theory in the Einstein representation,
another authors\cite{sakai,dalia} develop the formalism of a
singular hypersurface in the Jordan-Fierz (J-F) representation. We
shall follow the last approach in which it is usually considered
that the metric corresponds to the physical one.

In the weak field method of Barros and Romero if the metric field
equation is considered in the Einstein representation, it does not
depend on the scalar field. As a consequence,  the dynamics of the
singular hypersurface can be obtained in a similar manner as in
the corresponding model of GR.  Of course, the scalar field and
the metric in the JF representation  will depend on the matter
model assumed for the ejected shell, while   the radiation
involved in the process   will include two components, one
corresponding to the scalar field and the other to the
electromagnetic radiation, both tuned with the evolution of their
sources. Thus, once we have determined the evolution of the
sources, we know  the interior and exterior solutions and the
content and variation of the radiated energy at infinity in both
components.

In section 2 we present a brief summary of the results of the
model of Barros and Romero\cite{barros}. Section 3 is devoted to
present the results of the theory of surface layers in BD theory,
useful in the weak field limit. In section 4 we discuss a general
model of an stellar explosion by constructing the metric and the
scalar field both in the interior and exterior regions of the
singular hypersurface. We end the section with a derivation of the
frequency shift of spectral lines in the background metric and the
luminosity of the exploding object as seen by a distant observer
at rest. In section 5 we treat a particular model by matching  two
Vaidya  solutions, with time dependent masses $m^-$ and $m^+$,
respectively, through a shell of dust\cite{hs}, to obtain the
solution of Einstein equation, necessary to construct the BD
solution in its linearized form, for the same matter distribution.
The time evolution of the shell of dust is completely determined
by the GR matching conditions for the metric alone. The resulting
system of equations of motion consists of two ordinary
differential equations for four unknowns. To complete the system,
we obtain, first, an equation that shows how the total rest mass
of the shell changes due to the balance between the total energy
per unit time incident on the shell, $J_-$, and the total energy
per unit time emitted by the shell, $J^+$, both  at the shell
surface. Then, according to the general characteristic of the
scenario in which we are interested in, we may consider that $J^-
\ll J^+$, and therefore, $J^-$ it is neglected in the energy
balance equation. This assumption is equivalent to consider $m^-$
as constant. Another assumption is to take, essentially, $J^+$
proportional to the radius of the shell through a constant,
$\chi$, that it is characteristic of the radiation production
mechanism within the shell. This choice is similar  to that one
made by Hamity and Gleiser\cite{hamity2}, on different grounds,
that leads to a satisfactory model of a stellar explosion, with
characteristic parameters such as maximum luminosity and time
decay from that maximum, corresponding to a Type Ia
supernova\cite{supern}. This is precisely the scenario in which we
propose to compute the output of scalar energy from the ejected
shell.  For this particular example we compute the effective
active mass of the system as a function of an external observer
time. This effective mass presents two components: the ``tensor"
component, related to the radiation of electromagnetic energy, and
the scalar component, related to the radiation of scalar field. We
end the section with a numerical integration of the system of
differential equations from a given set of initial conditions. We
show a comparative set of results for different values of the
separation constant $\chi$, the Brans-Dicke parameter $\omega$,
and the initial active mass of the shell. One interesting result
is that the  active mass, associated with the scalar field, is
totally radiated to infinity; for the given system of initial
conditions the process takes place in a short time, well before
the star reaches its maximum luminosity if the initial active mass
of the shell is of the order of or greater than $10^{-2}M_\odot$.
This represents a mass loss in the ratio of the ``tensor"
component to the scalar component of 1 to $(2 \, \omega + 3)$, in
agreement with a general result of Hawking \cite{hawking}. Then,
this model shows explicitly, in a dynamical case, the mechanism of
radiation of scalar field, which is necessary to understand the
Hawking result. In the last section we summarize the main results
of our work.

\section{The weak field limit}

In the weak field approximation the solutions of BD equations are
simply related to the solutions of GR for the same matter
distribution. Following  Barros and Romero\cite{barros} the
corresponding field equations are:
\begin{eqnarray}
g_{ab} (x) &=& (G_0 \phi)^{-1} \bar{g}_{ab} (G_0, x) \; \;, \label{1}\\
g_{ab} (x) &=& \eta_{ab} + h_{ab} \;,\; h_{ab} \ll 1 \;\;, \label{1'}\\
\phi(x) &=& \phi_0 + \epsilon (x) \;\; ,\label{2}\\
\bar{G}_{ab} &=& 8 \pi G_0 T_{ab} \;\; ,\label{3}\\
\Box  \epsilon &=& \frac{8 \pi T}{2 \omega + 3}\;\; , \label{4}
\end{eqnarray}
The ``bar" on top of a symbol means that it is considered in the
Einstein representation. For instance, $\bar{g}_{ab} (G_0, x)$ is
the solution of eq.(\ref{3}). The same symbol without the bar is
in the J-F representation which it is usually considered as the
physical one. Thus, the physical metric is obtained from
eq.(\ref{1}).  The box operator $\Box$ corresponds to the wave
equation for the flat metric; $T_{ab}$ is the (matter) energy
momentum tensor. In first order $\bar{T}_{ab} = T_{ab}$; $T$ is
the trace of $T_{ab}$; $\epsilon (x)$ is a first order term in the
energy density; $\phi_0$ is an arbitrary constant which satisfies
$|\phi_0| \gg |\epsilon| $; $G_0 = \phi_0^{-1}$. Actually, in
order that BD theory posses a Newtonian limit this constant must
be related to the Newtonian gravitational constant $G$
by\cite{brans2}
\begin{equation}
G_0 = \left( \frac{2 \omega + 3}{2 \omega + 4}\right)\, G \;\;.
\label{2'}
\end{equation}
Essentially, in the weak field approximation the metric calculated
from BD equations is quasi-conformally related to the metric
calculated from Einstein equations for the same matter
configuration. The term quasi-conform means that in going from an
Einstein solution $\bar{g}_{ab} (G, x)$ to the corresponding BD
solution $g_{ab} (x)$, apart from the conformal factor $[1 - G_0
\, \epsilon(x)]$ we must replace $G$ by the new $\omega$-dependent
``effective" gravitational constant $G_0$ given by (\ref{2'}).

\section{Singular hypersurface in BD theory}

The procedure sketched in the previous section requires to find
first the metric $\bar{g}_{ab}$. To this end we have to specify
the tensor  $T_{ab}$ corresponding to the matter. We choose to
have a singular hypersurface of matter, $\Sigma$, that separates
the spacetime, $M$, into the external region $M^+$, with metric
$g^+_{ab}$, and the internal region $M^-$, with metric $g^-_{ab}$:
$M = M^+ \cup \Sigma \cup M^-$. The regions $M^+$ and $M^-$ are
non-empties; this a necessary requirement to show that the weak
field condition (\ref{1'})  guarantees the condition (\ref{2}) in
$M$.

Before we propose  the solutions $\bar{g}^{\pm}_{ab}$ and
$\epsilon^{\pm}$ in the regions $M^{\pm}$ we have to consider the
matching conditions at $\Sigma$ that such solutions have to
satisfy. Of course, the matching conditions of the metrics
$\bar{g}^+$ and $\bar{g}^-$ are the same as in GR\cite{Israel}:
\begin{equation}
8 \pi G_0 S = (\bar{K}^+ - \bar{K}^-\,) - \tilde{\bar{g}} \,tr(
\bar{K}^+ - \bar{K}^- )\;\;, \label{5}
\end{equation}
where $\bar{K}^+$ and $\bar{K}^-$ are the corresponding extrinsic
curvatures of $\Sigma$ as ``seen" from $M^+$ or $M^-$ in terms of
the metrics $\bar{g}^{\pm}_{ab}$, respectively; $\bar{K}^+$ and
$\bar{K}^-$ are tensors in  $\Sigma$; $\tilde{\bar{g}}$ is the
induced metric  on $\Sigma$ by the metric $\bar{g}$ of the
spacetime $M$. Finally, $S$ is the surface energy momentum tensor
in $\Sigma$.

The extrinsic curvature $\bar{K}^+$ can be expressed in terms of
its components in a coordinate basis $\{e_i\}$ in $\Sigma$:
\begin{equation}
\bar{K}^+_{ij} = e^a_i e^b_j \bar{K}^+_{ab}\;\;\;, \bar{K}^+_{ab}
= - (\left. h^c_a h^d_b \bar{N}_{c;d}\,)\right|_{\Sigma^+}
\;\;,\;\; h^c_a = \delta^c_a + \bar{N}_a \bar{N}^c \;\;, \label{6}
\end{equation}
where $\left.\bar{N}\right|_{\Sigma^+}$ is the unit normal vector
to $\Sigma$, pointing from $M^-$ to $M^+$, with its components
expressed in $M^+$; In eq. (\ref{6}) ``;" indicates the covariant
derivative using the Riemannian connection of $M^+$ associated
with $\bar{g}^{+}_{ab}$. Similar expressions correspond to
$\bar{K}^-_{ij}$.

On the other hand, eq.(\ref{4}) is in the  J-F representation. The
corresponding matching conditions on  the scalar field $\phi$
through $\Sigma$ are:
\begin{equation}
\left. \epsilon \right|_{\Sigma^+} = \left. \epsilon \right|_{\Sigma^-}\;\;, \label{7}
\end{equation}
while its normal derivative has a discontinuity given by\cite{dalia}:
\begin{equation}
[N(\epsilon)]_{\Sigma^+} - [N(\epsilon)]_{\Sigma^-} = -\, \frac{8 \pi \, tr S}{3 + 2\,\omega}\;\;, \label{8}
\end{equation}
where $tr S$ is the trace of  $S$,
\begin{equation}
[N(\epsilon)]_{\Sigma^+} = \left(N^a \frac{\partial \epsilon}{\partial x^a} \right)_{\Sigma^+} \;\; , \label{9}
\end{equation}
and a similar expression for ${\Sigma^-}$.

\section{The model}
\subsection{General characteristics of the model}

To construct a simple model of an stellar explosion  we choose
$\Sigma$ as the history of an spherical surface with metric
\begin{equation}
d s_\Sigma^2 = d \tau^2 - R^2 (\tau) d \Omega^2 \;\;. \label{10}
\end{equation}
Also, we consider $M^+$ filled with a coherent unpolarized radial
flow of electromagnetic radiation represented by
\begin{equation}
T^+_{ab} = (\rho \, k_a k_b)^+ \;\;, \label{11}
\end{equation}
where $k^+_a$ is a null vector. Then, the exterior metric solution
of eq.(\ref{3}) is represented by Vaidya's metric\cite{vaidya}:
\begin{equation}
d\bar{s}^{\,2}_+\, = \, \left( 1 - \frac{2 G_0 m^+(\mu)}{r}\right)
d \mu^2 + 2\, d \mu\,d r - r^2 d\Omega^2\;\;, \label{12}
\end{equation}
with
\begin{equation}
k^{+a} = \delta^a_r\;\;\;\;, \;\;\;\;\frac{d\, m^+}{d \mu} =  -
\,4 \pi r^2\,  \rho^+\;\;. \label{13}
\end{equation}
The scalar field in the exterior region satisfies eq.(\ref{4}) with $T = 0$. An outgoing wave solution is:
\begin{equation}
\epsilon^+ (x) = \frac{f (\mu)}{r}\;\;, \label{14}
\end{equation}
where $f (\mu)$ is an arbitrary function at the moment.

We consider a general model corresponding to:

\begin{itemize}
\item A time dependent spherical shell of matter with an energy momentum tensor,
compatible with the symmetry of the problem, given by
\begin{equation}
S =  \eta \, v \otimes v + p \,(v \otimes v - \tilde{g} ) \;\;
,\;\; \mbox{the basis  $\{e_j\} =  \{\partial / \partial \tau ,
\partial / \partial \theta, \partial / \partial \varphi \}\;\; ,$}
\label{15}
\end{equation}
where $v$ is the velocity of a fluid element in  $\Sigma$, $\eta$
the surface energy density and  $p$ the isotropic pressure within
the shell.
\item The metric $g^-_{ab}$ has the general form
\begin{equation}
ds^{\,2}_-\, = \, F^2 (\xi,r) d\xi^2 - H^2 (\xi,r) dr^2 - r^2
d\Omega^2\;\;, \label{16}
\end{equation}
where we have chosen the area variable $r$ as one of the coordinates, for simplicity.
\item The metric $g^+_{ab}$ in first order takes the form
\begin{equation}
ds^{\,2}_+\, = \, \left( 1 - \frac{2 G_0 m^+(\mu)}{r} - G_0
\,\epsilon^+ \right) d \mu^2 + \, (1 - \,G_0 \,\epsilon^+)(2\,d
\mu\,d r - \,r^2 d\Omega^2)\;. \label{16'}
\end{equation}

\item The interior of the shell is occupied by a central spherical body and, perhaps, electromagnetic
radiation and traceless  matter in a neighborhood of $\Sigma^-$,
such that in that region, close to $\Sigma$,  the scalar field
solution of eq.(\ref{4}) is
\begin{equation}
\epsilon^- (x) = \frac{g_1 (\xi + r)}{r} + \frac{g_2 (\xi - r)}{r}
\;\; ; \label{17}
\end{equation}
i.e., it is the sum of a spherical wave moving towards decreasing
$r$ plus a spherical wave travelling towards increasing  values of
$r$. The functions $g_1$ and $g_2$ may be determined by imposing
initial and boundary conditions in $M^-$ and on $\Sigma$.
\end{itemize}
From eqs.(\ref{5}), (\ref{7}) and (\ref{8}) we know that the
scalar field and the dynamics of the spherical shell are closely
related; that dynamics is contained in the solution of
eq.(\ref{3}). From eq.(\ref{7}) we obtain
\begin{equation}
f [ \mu (\tau)] = g_1[\xi(\tau) + R(\tau)] + g_2[\xi(\tau) -
R(\tau)]\;\;. \label{18}
\end{equation}
To apply the matching condition (\ref{8}) we need to know the
components $N^{\pm}_{a}$ in the J-F representation of the normal
vector $N$, with the requirement   $N^{\pm}_{r}> 0$. Those
components may be obtained from the normalization conditions:
\begin{equation}
v \cdot v = 1 \;\;, \;\; v \cdot N = 0 \;\; ,\;\; N \cdot N = - 1\;\;, \label{19}
\end{equation}
where the components $v^a = (X, \dot{R}, 0, 0)$ has to be computed
on both sides of $\Sigma$, separately. In particular on $\Sigma^-$
the time component  $X^- = d \xi(\tau)/d \tau$; similarly, on
$\Sigma^+$ is  $ X^+ = d \mu (\tau)/d \tau$. From (\ref{19}) we
obtain
\begin{equation}
N^{+ r} = - B\;\;,\; N^{+ \mu} =  X^+\;\;,\;N^{- r} = - (F/H)
X^-\;\;,\; N^{- \xi} = - (H/F) \dot{R} \;\;. \label{20}
\end{equation}
The definitions of the symbols used in (\ref{20}) are:
\begin{eqnarray}
B &=& + \,\left\{ \dot{R}^2 +  1 - \frac{G_0 }{R}(2 \,m^+ - f) \right\}^{1/2}\;\;, \label{21}\\
X^+ &=& (1 + \,G_0 \,\epsilon^+)\,(\dot{R} + B)^{-1} \;\; .
\label{22}
\end{eqnarray}
The expressions for $X^-$, $H(\xi, r)$, and $F(\xi, r)$ can only
be known symbolically and up to the first order in our
approximation. However, this knowledge is enough for the purpose
of the present work. In particular we have
\begin{equation}
X^- = 1 - (1/2)[ \dot{R}^2 - \varphi_{00} (R,\xi)]\;\; ,
\label{23}
\end{equation}
where $\varphi_{00} (\xi, r)$ is the first order term in the
expansion of the $(00)$ component of the metric. A straightforward
calculation in first order gives
\begin{eqnarray}
\left. N(\epsilon)\right|_{\Sigma^-} &=& - (X^- + \dot{R})\,
\frac{g\,'_{1}}{R} + (X^- - \dot{R}) \frac{g\,'_{2}}{R} +
\frac{g_1 + g_2}{R^2}\, X^- \;\;, \label{24}
  \\
\left. N(\epsilon)\right|_{\Sigma^+ }&=&  \frac{X^+ \,f'}{R} +
\frac{B\,f}{R^2}\;\;,  \label{25}
\end{eqnarray}
the (') indicates the derivative of the function. Computing the
derivative of expression (\ref{18}) with respect to $\tau$  we
obtain\begin{equation} g\,'_1 (X^- + \dot{R}) + g\,'_2 (X^- -
\dot{R}) - f'\,X^+ = 0 \;\; . \label{26}
\end{equation}
Therefore, from eqs.(\ref{24}), (\ref{25}) and (\ref{26}) we have
\begin{equation}
\left. N(\epsilon)\right|_{\Sigma^+} - \left.
N(\epsilon)\right|_{\Sigma^-} =  (X^- + \dot{R})\,\frac{2
g\,'_1}{R}  + (B - X^-) \, \frac{f}{R^2}\;\; . \label{27}
\end{equation}
Finally, taking into account that in first order $(B - X^-) \simeq
\dot{R}^2$, $g\,'_1 (X^- + \dot{R}) = d g_1/ d \tau$, and
eq.(\ref{8}) we have
\begin{equation}
\frac{d g_1}{d \tau} = - \, \frac{4 \, \pi \,R  (\eta - 2 p)}{3 +
2 \omega} - \frac{\dot{R}^2\, f}{2 R}\;\;. \label{28}
\end{equation}
Equation (\ref{28}) is valid in the first order of approximation,
including the terms in $\dot{R}^2$. This last equation, along with
eq. (\ref{5}), complete the description of a general model.

\subsection{Frequency shift and luminosity}

Consider\footnote{From now on we choose units such that $G = 1$;
i.e., $G_0 = (2\, \omega + 3)/ (2\, \omega + 4)$.} in $M^+$ an
observer at rest at spatial infinity and let $t$ measure proper
time along its world-line; the observer 4-velocity $u = (
\partial / \partial t)$ has components $u^{\mu} = 1$; $u^j =
0$ with $j = r, \theta, \phi$. Electromagnetic radiation emitted
at the surface of the shell with characteristic frequency
$\omega_e = k \cdot v (= X^+)$ will be received at spatial
infinity with frequency $\omega_r = k \cdot u (= 1)$. Thus,
\begin{equation}
\frac{\omega_e}{\omega_r} = \frac{1 + G_0\,\epsilon^+}{\dot{R} +
\left[ \dot{R}^2 +  1 - G_0\,\frac{2 \,m^+ - f}{R} \right]^{1/2}}
\equiv \frac{d t}{d \tau} ( = X^+ )\;\;. \label{35}
\end{equation}
We assume that $(2 \,m^+ - f) > 0$. Notice that $(\omega_e /
\omega_r) \rightarrow \infty$ (infinite redshift) for $R
\rightarrow \,G_0\,(2 \,m^+ - f)$ and $\dot{R} < 0$.

The total electromagnetic energy radiated by the shell in all
direction per unit time (the luminosity of the system) measured by
an observer at infinity is given by
\begin{equation}
L = \lim_{r\to\infty} 4 \pi \, r^2\, \left.(T_{ab} \tilde{u}^a
\tilde{N}^b)\right|_r = -\, \dot{m}^+/X^+ \;\; ,\label{36}
\end{equation}
where $\tilde{N}$ is the unit normal vector to the surface of
constant $r$ and $\tilde{u}$ the 4-velocity of an observer at rest
on that surface. The total electromagnetic energy ${\cal E}_{em}$
radiated in a time interval $\Delta t$ is given by
\begin{equation}
{{\cal E}_{em}} = \int_{\Delta t} L \, dt = m^+ (\tau) - m^+ (\tau
+ \Delta \tau)  \;\; \; \Rightarrow \;\;\; \dot{{\cal E}}_{em} =
-\, \dot{m}^+ \;\;,\label{37}
\end{equation}
In the case of a collapsing shell ($\dot{R} < 0$),  eq.(\ref{36})
shows that $L \rightarrow 0$ faster than $\dot{m}^+$ as $R
\rightarrow \,G_0\,(2 \,m^+ - f) $ due to an extra redshift
factor.

\section{A particular solution}

To present a particular simple solution of the general model  we
assume that:
\begin{itemize}
\item The metric $\bar{g}^-$ is also represented, in part, by a Vaidya solution,
corresponding to the exterior gravitational field of a radially radiating spherically
symmetric central body (the core of the exploding star), given by
\begin{equation}
d\bar{s}^{\,2}_-\, = \, \left( 1 - \,\frac{2\,
G_0\,m^-(\nu)}{r}\right) d \nu^2 + 2\, d \nu\,d r - r^2
d\Omega^2\;\;; \label{38}
\end{equation}
i.e, we also consider that in the interior of $M^-$, close to
$\Sigma$, we have a coherent unpolarized radial flow of
electromagnetic radiation represented by
\begin{equation}
T^-_{ab} = (\rho \, k_a k_b)^- \;\;, \label{11'}
\end{equation}
where $k^-_a$ is a null vector. with
\begin{equation}
k^{-a} = \delta^a_r\;\;\;\;, \;\;\;\;\frac{d\, m^-}{d \nu} =  -
\,4 \pi r^2\,  \rho^-\;\;. \label{13'}
\end{equation}
\item The hypersurface $\Sigma$ is the history of a spherical shell of dust:
\begin{equation}
 S = \eta \, v \otimes v \;\;. \label{38'}
\end{equation}
\end{itemize}
The equation of motion for the surface layer, as given by the
matching conditions eqs.(\ref{5}) and (\ref{6}) are
\begin{eqnarray}
R (\bar{A} - \bar{B}) =  4 \pi\,G_0\,  \eta R^2 &=:& m_0\;\;  \label{39} \\
\frac{1}{\bar{B}}\left(\ddot{R} + \frac{G_0\,m^+}{R^2} - \frac{
G_0\,\bar{X}^+\, \dot{m}^+}{R}\right) -
\frac{1}{\bar{A}}\left(\ddot{R} + \frac{G_0\,m^-}{R^2} - \frac{
G_0\,\bar{X}^-\, \dot{m}^-}{R} \right) &=& 4 \pi\,G_0\, \eta \;\;
, \label{40}
\end{eqnarray}
where
\begin{eqnarray}
\bar{X}^+ &=& (\dot{R} + \bar{B})^{-1} \;\; , \label{41}\\
\bar{X}^- &=& (\dot{R} + \bar{A})^{-1} \;\; , \label{41'}\\
\bar{B} &=& +
\,\left( \dot{R}^2 +  1 - \frac{2\,G_0\,  m^+}{R} \right)^{1/2}\;\;,\label{42}\\
 \bar{A} &=& + \,\left( \dot{R}^2 +  1 - \frac{2\,G_0\,
m^-}{R} \right)^{1/2}\;\;. \label{43}
\end{eqnarray}
We have introduced the total mass $m_0$ of the constituents if
infinitely dispersed and at rest\cite{Israel}. We assume that
\begin{equation}
R(\tau)\geq 2 \,G_0\,  m^+ \geq 2\,G_0\,  m^- \geq 0 \;\;.
\label{44}
\end{equation}
The momentum densities (or energy current densities) along the
unit normal vector $N$ to $\Sigma^\pm$ [$\Sigma^+$ (or $\Sigma^-$)
refers to $\Sigma$ when it is considered as part of $M^+$ (or
$M^-$)] measured by a local observer at rest on $\Sigma$
(4-velocity $v=
\partial/\partial \tau$) are given by
\begin{equation}
(T^{Nv})^\pm =: [\rho \, (k \cdot N)(k \cdot v)]^\pm = \rho^\pm
\,(\bar{X}^\pm)^2 \;\;. \label{43'}
\end{equation}
To write (\ref{43'})we have used the orthogonality conditions $N
\cdot v = 0$, $v \cdot v = 1$, $N \cdot N = - 1$,  where all the
scalar products are computed with the metric $\bar{g}$. The total
energy per time unit incident on ($J^-$), or emitted by ($J^+$),
the shell is
\begin{equation}
J^\pm =: 4 \pi R^2\, (T^{Nv})^\pm = -\,\dot{m}^\pm \bar{X}^\pm
\;\;. \label{44'}
\end{equation}
Taking the $\tau$-derivative of (\ref{39}) to use the resulting
equation in (\ref{40}), considering $\dot{R} \neq 0$, and the
definitions (\ref{41}-\ref{43}), we obtain\cite{hs}
\begin{equation}
\dot{m}_0 = J^- - J^+ \;\;, \label{45}
\end{equation}
which is an energy-balance equation.
It is easy to show that
equation (\ref{39}) and the definitions (\ref{42}) and (\ref{43})
give
\begin{equation}
\bar{A} = \frac{\hat{m}}{m_0} + \frac{m_o}{2R}\;,\;\; \bar{B} =
\frac{\hat{m}}{m_0} - \frac{m_o}{2R}\;,\;\;\hat{m} = m^+ -
m^-\;\;. \label{45'}
\end{equation}

If the functions $m^+ (\tau)$ and $m^- (\tau)$ are given,
(\ref{39}) and (\ref{40}) represent a system of second-order
ordinary differential equations for the unknowns $R(\tau)$ and
$\eta(\tau)$.

\subsection{The equation of motion as a first order system}

We are interested in a supernova explosion as the scenario to
study, in a specific example, the main characteristics of  scalar
radiation in the BD theory. To this end we shall make some further
assumptions to simplify the equation system. Let us consider first
eq. (\ref{45}). Once the gigantic explosion of the star takes
place its core may collapse so rapidly that it forms a sort of
extremely compact (degenerate) matter. This compact object, which
may be a neutron star\footnote{At the moment of a neutron star's
birth, the nucleons that compose it have energies characteristic
of free fall, which is to say about $100 MeV$ per nucleon. That
translates to $10^{12}\, K$ or so. The star cools off very
quickly, though, by neutrino emission, so that within a couple of
seconds the temperature is below $10^{11}\, K$ and falling fast.
In this early stage of a neutron star's life neutrinos are
produced copiously, and since if the neutrinos have energies less
than about $10 MeV$ they sail right through the neutron star and
the surrounding matter without interacting, they act as a
wonderful heat sink\cite{miller}.} or a black hole, is referred to
as a compact supernova remnant. It may also be present a diffuse
supernova remnant as a consequence of the shock wave and ejected
material expanding from this explosion, and the interstellar
material it sweeps up along the way. In our very crude model of
the explosion, part of the interior of the shell is represented by
a radiating compact object of mass $m^-(\tau)$ whose action on the
rest mass
 of the expanding shell is represented by $J^-$ in (\ref{45}). It
 seems then reasonable to assume\footnote{Since $J^-$ and $J^+$ are both functions of time, the
least favorable instant is at $\tau = 0$. In the next subsection
we shall see from the initial values of  all the relevant
variables that $J_i^+ > 10^4 J_\odot$.  } that $J^- \ll J^+$ and
neglect $J^-$ in (\ref{45}). This assumption is equivalent to
consider $m^-$ as a constant. Thus, eq. (\ref{45}) becomes
\begin{equation}
\dot{m}_0 = - J^+ (= - \,\dot{m}^+\, \bar{X}^+) \;\;. \label{47}
\end{equation}

In consonant with the last assumption we assume that the rate of
scalar radiation, that originates at $\Sigma$, to the interior and
exterior of the shell of dust, are equal; i.e.:
\begin{equation}
 \left. \frac{d g_1}{d \tau}\right|_{\Sigma_-} = \left. \frac{d f}{d \tau}\right|_{\Sigma_+}\;\;. \label{38"}
\end{equation}
According to (\ref{18}) what this assumption  actually means is
that $g_2$ is  constant; i.e., we have  in part of  $M^-$, in
first order, an incoming scalar wave that originates at $\Sigma$,
plus a static scalar field associated with the central body.

Let us introduce now the function $a(\tau)$ by
\begin{equation}
a(\tau)= \hat{m}/m_0 \;\;. \label{46}
\end{equation}
In terms of the parameter $a$ and using eqs. (\ref{42}),
(\ref{43}), and (\ref{45'}) we obtain a first order equation for
$R(\tau)$ in the form\cite{hamity2}
\begin{equation}
\dot{R}^2  = a^2 - 1 + \,G_0\,\frac{m^+ + m^-}{R} -
\frac{G_0^2\,\hat{m}^2}{4 a^2 R^2} \;\;. \label{49}
\end{equation}
Similarly, from (\ref{47}) and (\ref{46}) we have
\begin{equation}
\hat{m} \,\dot{a} = a \,\dot{\hat{m}}\,(1 - a\, \bar{X}^+)\;\;.
\label{51}
\end{equation}
Replacing first $\bar{B}$ given in (\ref{45'}) into (\ref{41}) and
then the resulting expression for $\bar{X}^+$ into (\ref{51}), we
obtain
\begin{equation}
\hat{m} \dot{a} (2 a R \dot{R} + 2 a^2 R - G_0 m_0) = a
\dot{\hat{m}} (2 a R \dot{R}  - G_0 m_0) \;\;. \label{51'}
\end{equation}
 Eqs. (\ref{49}) and (\ref{51'})   are two relations for the three
unknowns, $\hat{m}$, $a$ and $R$. Condition (\ref{44}) now reads
$a \ge 0$, $2 a^2 R \ge G_0\, \hat{m} \ge 0$.

To find an explicit solution it is necessary to fix one of the
unknown functions and then solve for the resulting differential
equations for the other two. To guide our intuition let us replace
the last obtained expression for $\bar{X}^+$ into (\ref{51}) and
then use (\ref{51'}) to replace $\dot{\hat{m}}$ in the result.
Thus, we obtain
\begin{equation}
\dot{a} = -\, \frac{J^+}{2 \hat{m} R}\,(2 a R \dot{R}  - G_0
m_0)\;\;. \label{52'}
\end{equation}
We assume now
\begin{equation}
J^+ = 2 \chi \hat{m}\,a\,R \;\;. \label{52"}
\end{equation}
The value of the constant $\chi$ is correlated to the time scale
of the process and it may be conjecture that it is a
characteristic of the radiation production mechanism within the
shell. For $\chi = 0$ we have a non-radiating system ($a$,
$\hat{m}$ = constants). The dimension of $\chi$ equals the inverse
of a square length; the adopted unit in this paper is
$M^{-2}_\odot$. This choice is equivalent to that one made by
Hamity and Gleiser\cite{hamity2}, on different grounds,  that
leads to a satisfactory model of a stellar explosion, with
characteristic parameters such as maximum luminosity and time
decay from that maximum, corresponding to a Type Ia supernova.
This is precisely the scenario in which we propose to compute the
output of scalar energy from the ejected shell\footnote{If we
assume that the total electromagnetic energy per unit time,
radiated by the shell, is proportional to its area times the
absolute temperature $T$ at its surface to the fourth power, we
have that $T^4 \sim \hat{m}\, a / R$, which is a decreasing
function of time during the expansion period.}. Finally, from
(\ref{52'}) and (\ref{52"}) we obtain
\begin{equation}
\dot{a} = - \chi \,a\, ( 2 a R \dot {R} - G_0\, \hat{m} )\;\;,\;\;
\chi = \mbox{const.} \geq 0  \;\;. \label{52}
\end{equation}
Hence, we can write  (\ref{49}) and (\ref{51'}) in the form
\begin{eqnarray}
\dot{R} &=& \pm \left( a^2 - 1 + G_0\,\frac{\hat{m} + 2\,m^-}{R} +
\frac{G_0^2 \,\hat{m}^2}{4 a^2 R^2}\right)^{1/2}\;\;, \label{49'}\\
\dot{\hat{m}} &=& -\; \chi \hat{m} \,(2 \,a \,R \,\dot{R} + 2 a^2
R - G_0 \,\hat{m}) \;\;. \label{60}
\end{eqnarray}

The system of first order differential equations, (\ref{52},
\ref{49'}, \ref{60}), that describes the motion of the shell, has
to be considered in conjunction with the first order differential
equation for the amplitude  of the outgoing scalar  wave, which
according to eq.(\ref{28}), and assumptions (\ref{38'}) and
(\ref{38"}), becomes
\begin{eqnarray}
\dot{f} = \dot{g}_1 &=& - \, \frac{4 \, \pi\, R^2  \eta }{R (3 + 2 \omega)} - \frac{\dot{R}^2 f}{2 R}\nonumber \\
&=& - \,\frac{\hat{m}}{a\,R \,(3 + 2 \omega)} - \frac{\dot{R}^2
f}{2 R}\;\;. \label{55'}
\end{eqnarray}
The initial condition for $f(\tau)$ may be obtained from the
requirement that at the time of the explosion, $\tau = 0$, the
exterior scalar field matches continuously to a BD static solution
in the weak field approximation generated by a central body of
active mass $m^+_i$\cite{barros}; i.e.,
\begin{equation}
f_i = \frac{2\,m_i^+}{(2\, \omega + 3)}\;\;, \label{57}
\end{equation}
where $f_i$ and $m^+_i$ are the initial values of $f(\tau)$ and
$m^+(\tau)$ respectively.

In order to compare the results of our model, in the
electromagnetic mode, with observational data  for a stellar
explosion,  such as a Type Ia Supernova\cite{kriscimas}, for
instance, we need to compute the light curve, $L(\tau)$,  from eq.
(\ref{36}), for values of $0 \leq \tau \leq \tau_{1}$, where
$\tau_{1}$ corresponds to the time when the maximum luminosity has
decline in one magnitude.\footnote{Actually, the maximum
luminosity should correspond to the B-band\cite{phillips} but in
our very crude model of a stellar explosion we do not have the
possibility to specify the B-band from the model of radial
electromagnetic radiation corresponding to the Vaidya solution.}
From eqs. (\ref{21}), (\ref{22}), (\ref{36}) and (\ref{49'}) the
expression for the light curve becomes
\begin{equation}
L(\tau) = -\, \dot{ \hat{m}} \left[\dot{R} + \left(a^2 +
\frac{G_0^2\,\hat{m}^2}{4\, a^2\,R^2} - \frac{G_0\, \hat{m}}{R} +
\frac{G_0\, f}{R}\right)^{1/2}\right]\left(1 + \frac{G_0\,f}{R}
\right)^{-1} \;\;. \label{36'}
\end{equation}

It is apparent  from eq.(\ref{16'}) that the Keplerian mass $M$
(the active mass),  measured by an orbiting object around the
exploding star, in the present approximation is
\begin{equation}
M = m^+ + f/2 \;\; ; \label{53}
\end{equation}
i.e., the total active mass is decomposed  into the sum of a
``tensor" component $m^+$ and a scalar  component $f/2$\cite{bh}.
Therefore, from eq.(\ref{16'}) the total scalar energy radiated in
the time interval $\Delta \tau$ is given by
\begin{equation}
\Delta\, {\cal{E}}_\phi = [f(\tau) - f(\tau + \Delta \tau)]/2
\;\;\; \Rightarrow\;\;\; \dot{{\cal E}}_\phi = - \frac{1}{2}
\dot{f}\;\;. \label{54}
\end{equation}
Thus, from eqs. (\ref{55'}) and (\ref{54}) the total active mass
radiated in the time interval $[0,\tau_1]$, corresponding to the
observer time interval  $[0,t_1]$, associated with the scalar
field is
\begin{equation}
{{\cal E}}_{\phi1} = \int_{\tau_1} \left[\frac{\hat{m}}{2\,a\,R(3
+ 2 \omega)} + \frac{\dot{R}^2 f}{4 R} \right]d\tau \;\;.
\label{56}
\end{equation}

\subsection{Numerical results}

We have performed a numerical integration of the system
(\ref{35},\ref{37},\ref{52},\ref{49'},\ref{60},\ref{55'},\ref{54}),
corresponding to the functions $t(\tau), \,{{\cal
E}}_{em}(\tau),\,a(\tau),\,R(\tau),\,\hat{m}(\tau),\,f(\tau)$, and
${{\cal E}}_{\phi}(\tau)$, with initial conditions, at $\tau= 0$:
$$t_i = 0,\, {{\cal
E}}_{emi} = 0,\, a_i^2 - 1 = 10^{-3},\, R_i = 2 \times 10^{-2}
R_\odot \equiv 9469.4 M_{\odot},\,$$
$$ \hat{m}_i = (10^{-3}, \,10^{-2},\,10^{-1}) \,M_{\odot}, \,f_i = 2\,m_i^+ /(2\, \omega +
3),\, {{\cal E}}_{\phi i}= 0.$$ The constant $\, m^- = 1
M_{\odot}$, and $\chi$ was chosen to have three different values,
corresponding to different cases. The initial value of the total
energy per unit time emitted by the shell verifies\footnote{The
corresponding value for a star like the Sun is $J^+ \sim 10^{33}
ergs/s$.} $J_i^+ > 3.5 \times 10^{37} ergs/s$. Similarly, for
these initial values and $\omega = 500$, the initial velocity of
the shell becomes $\dot{R}_i \approx 10000 \,km/s$. The value of
$\omega = 500$ it is generally accepted\cite{carroll} as the
lowest bound implies from Solar system tests, although some more
recent estimates\cite{carroll2} have raised this limit to $\omega
> 40000$, obtained using signal timing from the Cassini
spacecraft\cite{bertotti}. The same authors\cite{carroll2} point
out that this bound may be weaker on cosmological  scales than in
the solar system\cite{clifton}. On the other hand, in a recent
publication\cite{kim} it is shown that the parameter $\omega$
``runs" with the scale factor in a Friedman-Robertson-Walker
metric in order that a BD theory serves as a successful model for
dark matter - dark energy. In this model the value of the
parameter $\omega$ is less than $(-3/2)$ in the matter to scalar
transition period; equal to $(-3/2)$ in the BD scalar field
dominated era; and greater than $(3/2)$ in the scalar to
acceleration transition period. In any case, in our BD model of an
stellar explosion in the weak field approximation, the main
characteristics of the model are quite insensitive to any value of
$\omega$ greater than 2 since it results $G_0 \approx 1$, except
for {\sl the radiation rate of scalar field and the total energy
radiated in the scalar mode} as it can be seen from eqs.
(\ref{57}), (\ref{54}), and (\ref{56}).

For the given set of initial conditions we have computed the light
curve $L(\tau)$, for $0 \leq \tau \leq \tau_{1}$, and the total
radiated active masses ${{\cal E}}_{em1}$ and  ${{\cal E}}_{\phi
1}$ in the same time interval. In Table \ref{numeros} we show a
comparative set of results for different values of the constants
$\chi$ and $\omega$, and the initial values of $\hat{m}_i$;
$\triangle t_1 = t_m - t_1$, where $t_m$ and $t_1$ are the
observer times at the maximum luminosity, $L_m$, and when it has
decline in one magnitude, $L_1$, respectively\footnote{($L_m/L_1)
= 10^{2/5}$\cite{gaposchkin}.}. By $f_1$, we indicate the final
value of $f$ at observer time $t_1$. When $\hat{m}_i = 10^{-2}
M_\odot$, $f(\tau)$ becomes equal to zero at a time $t_f \approx
716 s$; while if $\hat{m}_i = 10^{-1} M_\odot$, it results $t_f
\approx 1.5 s$; in both cases $t_f \ll t_0$, where $t_0$ is the
observer time at which $\dot{R} = 0$; in all the cases $t_m <
t_0$. The reason for this fast approaching to zero of $f(\tau)$ is
that $\dot{f}$ is essentially proportional to $\hat{m}$, through
the first term on the right hand side of eq. (\ref{55'}). Finally,
it may be important to mention that although we have worked with
the exact Vaidya solution, for simplicity, the same numerical
results are obtained if we drop in the equations all the terms
quadratic in $\hat{m}$.

\begin{table}
\begin{center}
\begin{tabular}{||c|c|c|c|c|c|c|c||}
\hline $\chi$& $\omega$ & $\hat{m}_i$ &$L_m$ &$\triangle t_1$&$f_1$&${{\cal E}}_{em1}$ & ${{\cal E}}_{\phi1}$\\
\scriptsize{[$M_\odot]^{-2}$}&&\scriptsize{$[M_\odot]$}&\scriptsize{[ergs/s]}&\scriptsize{days}&\scriptsize{$[M_\odot]$}
&\scriptsize{[ergs]}&\scriptsize{[ergs]}\\
\hline
\footnotesize{$5\times10^{-24}$}&\footnotesize{500}&\footnotesize{$10^{-3}$}&\footnotesize{$3.53\times10^{43}$}&\footnotesize{22.1}&
\footnotesize{$15\times10^{-4}$}&\footnotesize{$1.04\times10^{50}$}&\footnotesize{$4.46\times10^{50}$}\\
\hline
\footnotesize{$10^{-23}$}&\footnotesize{500}&\footnotesize{$10^{-3}$}&\footnotesize{$4.98\times10^{43}$}&\footnotesize{15.8}
&\footnotesize{$15\times10^{-4}$}&\footnotesize{$1.04\times10^{50}$}&\footnotesize{$4.37\times10^{50}$}\\
\hline
\footnotesize{$25\times10^{-24}$}&\footnotesize{500}&\footnotesize{$10^{-3}$}&\footnotesize{$7.89\times10^{43}$}&\footnotesize{9.70}
&\footnotesize{$15\times10^{-4}$}&\footnotesize{$1.04\times10^{50}$}&\footnotesize{$4.22\times10^{50}$}\\
\hline
\footnotesize{$10^{-23}$}&\footnotesize{5000}&\footnotesize{$10^{-3}$}&\footnotesize{$4.98\times10^{43}$}&\footnotesize{15.7}
&\footnotesize{$15\times10^{-5}$}&\footnotesize{$1.04\times10^{50}$}&\footnotesize{$4.38\times10^{49}$}\\
\hline
\footnotesize{$10^{-23}$}&\footnotesize{50}&\footnotesize{$10^{-3}$}&\footnotesize{$4.98\times10^{43}$}&\footnotesize{15.7}
&\footnotesize{$14.7\times10^{-3}$}&\footnotesize{$1.04\times10^{50}$}&\footnotesize{$4.26\times10^{51}$}\\
\hline
\footnotesize{$10^{-23}$}&\footnotesize{5}&\footnotesize{$10^{-3}$}&\footnotesize{$4.98\times10^{43}$}&\footnotesize{15.6}
&\footnotesize{$12.0\times10^{-2}$}&\footnotesize{$1.04\times10^{50}$}&\footnotesize{$3.04\times10^{52}$}\\
\hline
\footnotesize{$10^{-23}$}&\footnotesize{500}&\footnotesize{$10^{-2}$}&\footnotesize{$4.99\times10^{44}$}&\footnotesize{15.8}
&\footnotesize{$^{(1)}$}&\footnotesize{$1.04\times10^{51}$}&\footnotesize{$^{(1)}$}\\
\hline
\footnotesize{$10^{-23}$}&\footnotesize{$5\times10^{4}$}&\footnotesize{$10^{-2}$}&\footnotesize{$4.99\times10^{44}$}&\footnotesize{15.8}
&\footnotesize{$^{(1)}$}&\footnotesize{$1.04\times10^{51}$}&\footnotesize{$^{(1)}$}\\
\hline
\footnotesize{$5\times10^{-24}$}&\footnotesize{$500$}&\footnotesize{$10^{-2}$}&\footnotesize{$3.53\times10^{44}$}&\footnotesize{21.9}
&\footnotesize{$^{(1)}$}&\footnotesize{$1.04\times10^{51}$}&\footnotesize{$^{(1)}$}\\
\hline
\footnotesize{$10^{-23}$}&\footnotesize{$500$}&\footnotesize{$10^{-1}$}&\footnotesize{$4.99\times10^{45}$}&\footnotesize{15.6}
&\footnotesize{$^{(1)}$}&\footnotesize{$1.04\times10^{52}$}&\footnotesize{$^{(1)}$}\\
\hline \multicolumn{8}{l}{$^{(1)}$\footnotesize{The active mass
associated with
the scalar field, $f_i/2$, is totally radiated to infinity after a time}} \\
\multicolumn{8}{l}{\footnotesize{  $t_f < t_1$, well before the
shell reaches its maximum brightness. For the values of $t_f$, see
the text.}}
\end{tabular}
\caption{A comparative set of results for different values of the
constants $\chi$ and $\omega$, and  $\hat{m}_i$; the other
parameters and $m^- = 1 M_\odot$, are the same in all cases.}
\label{numeros}
\end{center}
\end{table}

\section{Final comments}

We have treated a very crude model of an exploding star in the
weak field limit of BD theory. However, the model presents aspects
in the electromagnetic energy radiated to infinity that resembles
some characteristics data of a Type Ia Supernova. The most
noticeable feature is shown in the first three lines of Table
\ref{numeros}; i.e., the value of the constant $\chi$ relates the
absolute magnitude at maximum brightness and the decline rate in
one magnitude from that maximum. This characteristic has become
one of the most accurate method to measure luminosity distances to
objects at cosmological distances\cite{phillips, supern}. We also
notice from Table 1, that we may scale the value of $L_m$ by
changing just the initial value $\hat{m}_i$, without affecting the
decline rate $\triangle t_1$. These values seem to be independent
of the value of $\omega$ within a wide range. Finally, the total
electromagnetic energy, radiated in the time interval $[0,t_1]$,
is independent of the value of $\chi$ and $\omega$; the total
energy ${{\cal E}}_{\phi1}$, in those cases in which it is not
limited by the initial value $f_i$, it is larger the smaller the
value of $\omega$.

On the other hand, the total active mass associated with the scalar field, as given by $f_i/2$, is totally radiated to infinity. This process  takes place in a time lapse considerably smaller than the time in which the star reaches its maximum brightness after explosion, if the mass of the shell is of the order of or greater than $10^{-2}M_\odot$. This represents a mass loss in the ratio of the ``tensor" component to the scalar component of 1 to $(2\, \omega + 3)$, in agreement with a general result of Hawking\cite{hawking}.

\vspace{1cm}
\noindent \textbf{Acknowledgments}

The authors are very grateful to CONICET of Argentina, and SCyT of
the Universidad Nacional de C\'{o}rdoba for financial support.


\begin{thebibliography}{00}
\bibitem{phillips}M.M. Phillips, Ap.  J. {\bf 413}, L105 (1993).
\bibitem{supern}See: www.all-science-fair-projects.com/science\underline{
}fair\underline{ }projects\underline{ }encyclopedia/Supernova, for
a brief  description of Supernovae types
\bibitem{hawking}S.W. Hawking, Commn. math. Phys. {\bf 25}, 167
(1972).
\bibitem{vaidya}P.C. Vaidya, Indian Acad. Sci. {\bf A33}, 264 (1951); Nature {\bf 171}, 260 (1953).
\bibitem{antecedentes}W.J. Kaufmann III, J. Math. Phys. {\bf 9}, 1053 (1968).
\bibitem{brans}C.Brans, Ph.D. thesis, Princeton University (1961).
\bibitem{hamity2}V.H. Hamity and R.J. Gleiser, Astrophysics and Space Science {\bf 58}, 353 (1978).
\bibitem{lake}See, for instance:  K. Lake, Phys. Rev. D {\bf 19}, 2847 (1979).
\bibitem{hs} V.H. Hamity and R. Spinosa, J. of General Rel. and Grav. {\bf
16}, 9 (1984).
\bibitem{barros}A. Barros and C. Romero, Phys. Lett. {\bf A245}, 31 (1998).
\bibitem{sakai}Nobuyuki Sakai and Kei-ichi Maeda, Phys. Rev, D {\bf 48}, 5570 (1993).
\bibitem{dalia}D.S. Goldwirth and H.W. Zaglauer, Class. Quantum Grav. \textbf{10}, 1507 (1993).
\bibitem{barrabes}C. Barrab\`es and G.F. Bressange, Class. Quantum Grav.  \textbf{14}, 805 (1997).
\bibitem{brans2}C. Brans and R.H. Dicke, Phys. Rev. {\bf 124}, 925
(1961).
\bibitem{Israel}W. Israel, Nuovo Cimento {\bf B44S10},1 (1966);
Erratum-Ibid.{\bf 48B}, 463 (1967); Il Nuovo Cimento {\bf 44B}, 1
(1966).
\bibitem{miller} For a quick reference to neutorn stars see:
http://www.astro.umd.edu/$\sim$miller/nstar.html
\bibitem{kriscimas}K. Kriscimas \textit{et al}., ``Optical and Infrared
Photometry of Type Ia Supernovae 1991T, 1991bg, 1999ek, 2001bt,
2001cn, 2001cz, and 2002bo", arXiv: astro-ph/0409036 v1 (2 Sep
2004).
\bibitem{bh}D. Barraco and V. Hamity, Class. Quantum Grav. {\bf
11}, 2113 (1994).
\bibitem{carroll}See, for example, S.M. Carroll, \emph{Spacetime and Geometry}, (Addison Wesley, San Francisco, 2004).
\bibitem{carroll2}S.M. Carroll \textit{et al.}, arXiv: astro-ph/0408081 v2 (28 Sep 2004).
\bibitem{bertotti}B. Bertotti, L. Iess and P. Tortora, Nature,
\textbf{425}, 374 (2003).
\bibitem{clifton}T. Clifton, D.F. Mota and J.D. Barrow,
arXiv:gr-qc/0406001 v1 (1 Jun 2004).
\bibitem{kim}H. Kim, ``Brans-Dicke Theory as an Unified Model for
Dark Matter-Dark Energy", arXiv: astro-ph/0408577 v3 (10 Sep
2004).
\bibitem{gaposchkin}See, for instance, C. Payne-Gaposchkin
\emph{Introduction to Astronomy}, Chapter XII, (Universitry
Paperbacks, Methuen and Co., London, 1961)
\end{thebibliography}
\end{document}